\documentclass[conference]{IEEEtran}
\IEEEoverridecommandlockouts

\usepackage[dvipdfmx]{graphicx}
\usepackage{url}

\usepackage{multirow}
\usepackage{colortbl}
\usepackage{cite}
\usepackage{amsmath,amssymb,amsfonts}
\usepackage{algorithmic}
\usepackage{graphicx}
\usepackage{textcomp}
\usepackage{xcolor}
\def\BibTeX{{\rm B\kern-.05em{\sc i\kern-.025em b}\kern-.08em
    T\kern-.1667em\lower.7ex\hbox{E}\kern-.125emX}}
\begin{document}

\title{Comparing effects of price limit and circuit breaker \\ in stock exchanges by an agent-based model
\thanks{Note that the opinions contained herein are solely those of the authors and do not necessarily reflect those of SPARX Asset Management Co., Ltd.}
}

\author{\IEEEauthorblockN{Takanobu Mizuta}
\IEEEauthorblockA{\textit{SPARX Asset Management Co. Ltd.} \\
Tokyo, Japan \\
https://orcid.org/0000-0003-4329-0645}
\and
\IEEEauthorblockN{Isao Yagi}
\IEEEauthorblockA{\textit{Faculty of Informatics} \\
\textit{Kogakuin University}\\
Tokyo, Japan \\
https://orcid.org/0000-0003-0119-1366}}


\maketitle

\begin{abstract}
The prevention of rapidly and steeply falling market prices is vital to avoid financial crisis. To this end, some stock exchanges implement a price limit or a circuit breaker, and there has been intensive investigation into which regulation best prevents rapid and large variations in price. In this study, we examine this question using an artificial market model that is an agent-based model for a financial market. Our findings show that the price limit and the circuit breaker basically have the same effect when the parameters, limit price range and limit time range, are the same. However, the price limit is less effective when limit the time range is smaller than the cancel time range. With the price limit, many sell orders are accumulated around the lower limit price, and when the lower limit price is changed before the accumulated sell orders are cancelled, it leads to the accumulation of  sell orders of various prices. These accumulated sell orders essentially act as a wall against buy orders, thereby preventing price from rising. Caution should be taken in the sense that these results pertain to a limited situation. Specifically, our finding that the circuit breaker is better than the price limit should be adapted only in cases where the reason for falling prices is erroneous orders and when individual stocks are regulated. 
\end{abstract}


\section{Introduction} 
\label{s1}
The prevention of rapidly and steeply falling market prices is vital to avoid financial crisis. To this end, some stock exchanges implement a price limit, which refuses orders that are priced significantly away from the current market price, or implement a circuit breaker, which halts placing orders for a while when the market prices are largely varied. Stock exchanges in Japan, South Korea, China, and other Asian countries tend to implement the price limit, while those in the USA and Europe lean more toward the circuit breaker. While there has been intensive investigation into which regulation best prevents rapid and large variations in price\cite{KIM2008197}, this question remains unanswered. 

Empirical studies cannot be conducted to isolate the direct effect of implementing a trading regulation due to the many diverse factors affecting price formation in actual markets. In contrast, an artificial market model, which is an agent-based model for a financial market\footnote{The  basic concept for constructing and validating the artificial market model utilized in the current work are explained in Appendix \ref{s5.2} and \ref{s5.3}, or the article\cite{mizuta2019arxiv, mizuta2022aruka}}, can isolate the pure contributions of a trading regulation. Many previous artificial market studies have contributed to explaining the nature of financial market phenomena such as bubbles and crashes. Recent artificial market studies have also contributed to discussions about appropriate financial regulations and rules \cite{mizuta2016SSRNrev, mizuta2019arxiv, mizuta2022aruka}. The JPX\footnote{The Japan Exchange Group (JPX) is a Japanese financial services corporation operating the Tokyo Stock Exchange.} Working Paper series includes various studies that have contributed to such discussions\footnote{https://www.jpx.co.jp/english/corporate/research-study/working-paper/index.html}.

There have been many investigations using artificial market models for both the price limit\cite{yeh2010examining, mizuta2016ISAFM, zhang2016price} and the circuit breaker\cite{Muranaga1999halt, kobayashi2011benefits}. Mizuta et al.\cite{mizuta2016ISAFM} investigated the price limit and showed that the following conditions prevent large price variations:
\begin{equation}
Pr/tr<S_{fall}, \label{eqj1}
\end{equation}
\begin{equation}
Pr>Vol_{tr}, \label{eqj2}
\end{equation}
\begin{equation}
tr<tm, \label{eqj3}
\end{equation}
\begin{equation}
Pr<P_{DD}, \label{eqj4}
\end{equation}
where $Pr$ and $tr$ are parameters of the price limit (respectively a limit price range and a limit time range), $S_{fall}$ is the falling speed of market prices, $Vol_{tr}$ is the standard deviation of every $tr$ return (volatility), $t_{m}$ is the erroneous orders period, and $P_{DD}$ is the falling depth of market prices. Their findings showed that the $Pr$ and $tr$ parameters should be satisfied and that stock exchanges should implement several price limits for every time scale of price variety.

However, no previous research has compared the effectiveness of the price limit versus  the circuit breaker using the $Pr$ and $tr$ parameters, mostly because the model needs to be improved before the price limit and circuit breaker can be compared under an equal condition. For example, when a circuit breaker is active, time simply passes without orders being placed, but Mizuta et al.'s model\cite{mizuta2016ISAFM} cannot treat such time passing without placing orders because the time passes  only when an order is placed. 

Therefore, in this study, we expanded on Mizuta et al.'s artificial market model\cite{mizuta2016ISAFM} by adding stop loss behavior of the agents and a circuit breaker and then used it to investigate whether the price limit or the circuit breaker is more effective to prevent falling market prices. Each agent estimates a fair price and then re-estimates it when market prices fall significantly below that price. The agents need a long time for the re-estimation, so they place stop-loss orders during the re-estimation, which enables the model to let time pass without placing orders.

\begin{figure}[t] 
\begin{center}
\includegraphics[scale=0.30]{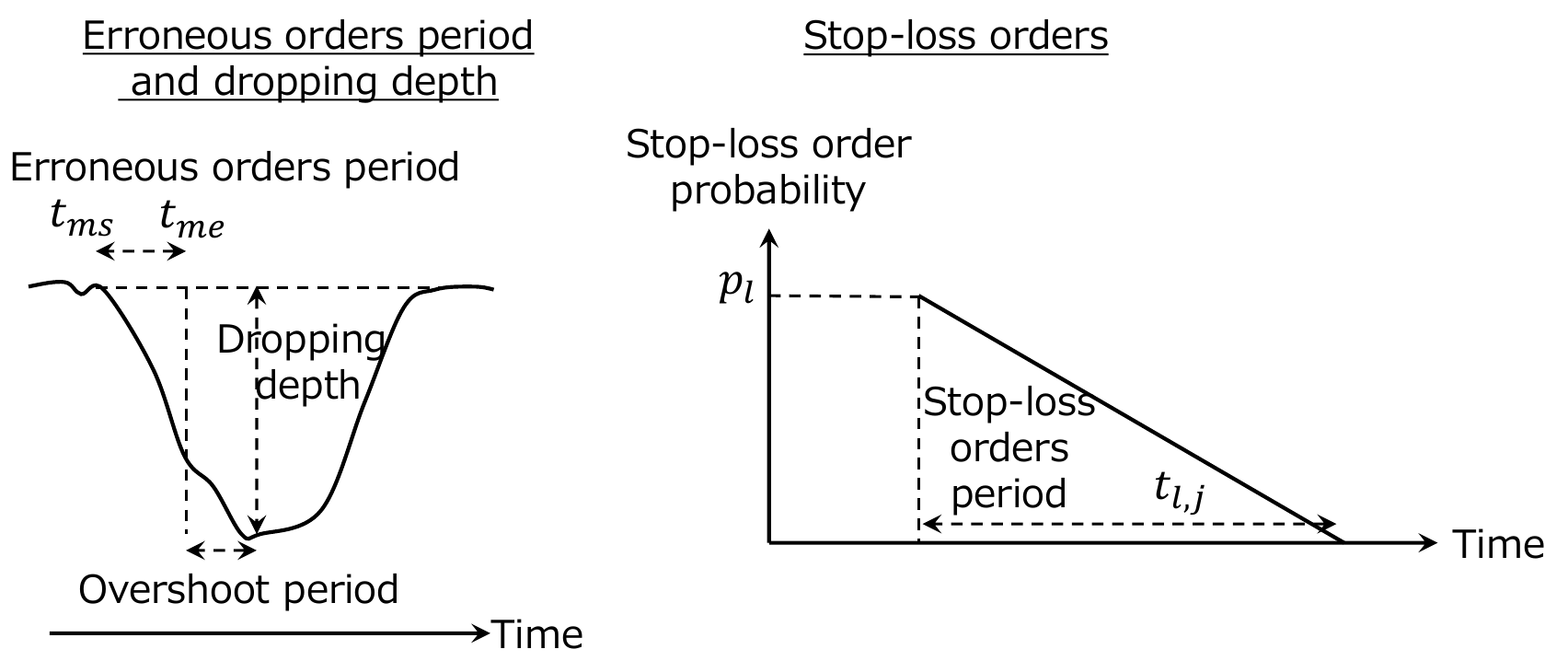}
\end{center}
\caption{Illustrations for (left) erroneous orders period and falling depth and (right) stop-loss orders. }
\label{p01}
\end{figure}

\begin{figure}[t] 
\begin{center}
\includegraphics[scale=0.25]{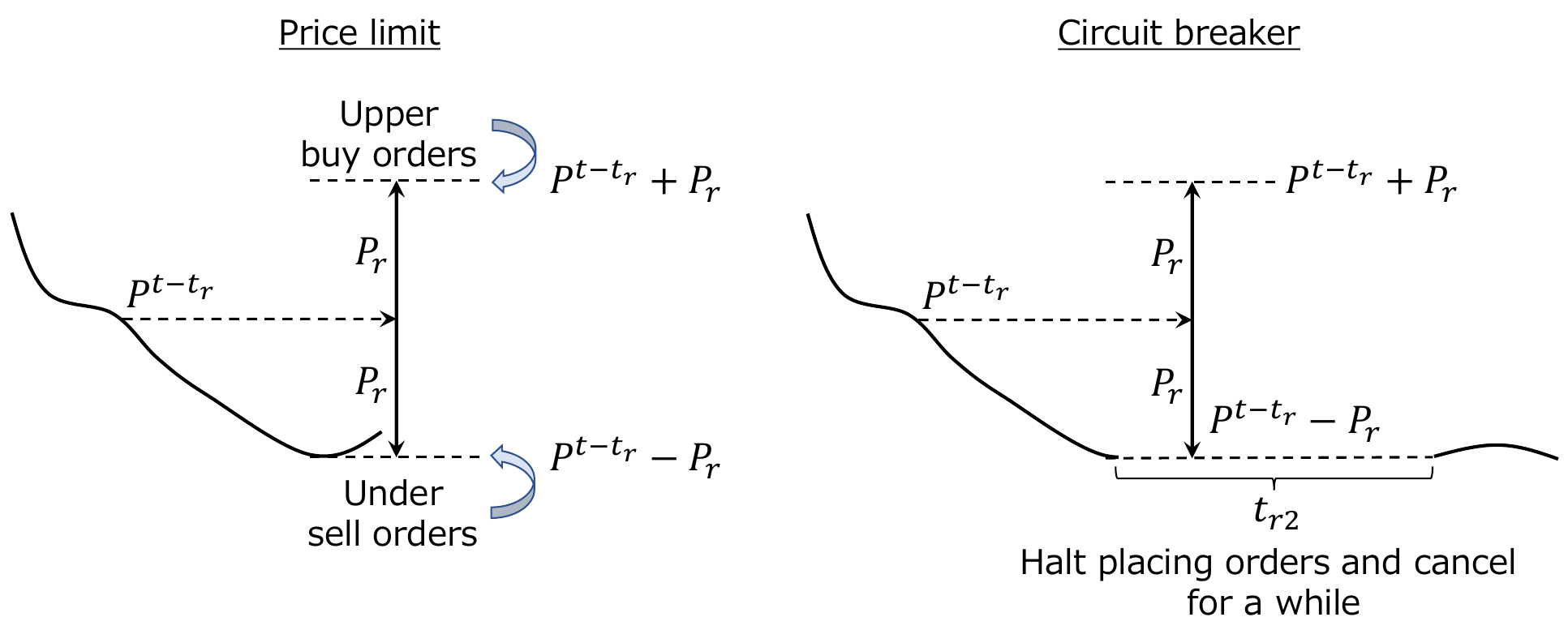}
\end{center}
\caption{Illustrations for (left) price limit and (right) circuit breaker. }
\label{p02}
\end{figure}

\section{Model}
\label{s3}
Chiarella and Iori\cite{chiarella2002simulation} built a model that, while very simple, could replicate long-term statistical characteristics observed in actual financial markets, e.g. fat-tail and volatility clustering. Mizuta et al.\cite{mizuta2016ISAFM} later expanded this model so that it can treat large fluctuation of prices such as turmoil by erroneous orders. Only fundamental and technical analysis strategies that exist generally for any market at any time\footnote{Many questionnaire-based empirical studies have found these strategies to be the majority of investment strategies, as comprehensively reviewed by Menkhoff and Taylor\cite{menkhoff2007}. The empirical study using market data by Yamamoto\cite{yamamotopredictor} showed that investors are switching fundamental and technical analysis strategies.} are implemented in the agent model.

The simplicity of the model is very important for this study because unnecessary replication of macro phenomena leads to models that are overfitted and too complex. Such models hamper the understanding and discovery of the mechanisms affecting price formation because of the increase in related factors. Gilbert said "the aim of abstract models is to demonstrate some basic social process that may lie behind many area of social life"\cite{gilbert2008agent}. Axelrod mentioned that to understand mechanism abstract models should be as simple as possible because needless complex model preventing to understand the mechanism, and called the principle as KISS (keep it simple stupid)\cite{10.2307/j.ctt7s951}. Because our model is abstract model to understand and discover the mechanisms mechanism, our model should obey KISS principle. The detail of basic concept for constructing our artificial market model and an explanation of its validation are explained in Appendix \ref{s5.2} and \ref{s5.3}, or the article\cite{mizuta2019arxiv, mizuta2022aruka}.

In the current study, we extend Mizuta et al.'s artificial market model\cite{mizuta2016ISAFM} by adding stop loss behavior of agents and a circuit breaker.

This model contains one stock, and the stock exchange utilizes a continuous double auction to determine the market price\cite{tse2019e}. In the auction mechanism, multiple buyers and sellers compete to buy and sell stocks in the market, and transactions can occur at any time an offer to buy and an offer to sell match. The minimum unit of a price change is $\delta P$. The buy-order and sell-order prices are respectively rounded down and up to the nearest fraction.

\subsection{Agents}
We introduce agents for modeling a general investor so as to replicate the nature of price formation in actual financial markets. The number of agents is $n$ and they can short sell freely. The holding positions are not limited, so agents can take an infinite number of shares for both long and short positions. Time $t$ increases by one when an agent places an order or when a circuit breaker in operation skips placing an order by an agent. 

Agents always order only one share at a time. First, agent $1$ places an order to buy or sell a stock, and then agent $2$ places an order to buy or sell. After that, agents $3,4,,,n$ each place orders to buy or sell. After the final agent $n$ places an order, going back to the first agent, agent $1$ places an order to buy or sell, and at agents $2,3,,,,n$ each place orders to buy or sell, and this cycle is repeated.

An agent determines the order price and buys or sells using a combination of fundamental and technical analysis strategies to form an expectation of the stock return. The expected return of agent $j$ at $t$ is calculated as
\begin{equation}
r^{t}_{e,j} = (w_{1,j} \ln{\frac{P_f}{P^{t-1}}} + w_{2,j}\ln{\frac{P^{t-1}}{P^{t-\tau _ j-1}}}+w_{3,j} \epsilon ^t _j )/\Sigma_i^3 w_{i,j}, \label{eq1}
\end{equation}
where $w_{i,j}$ is the weight of term $i$ for agent $j$ and is independently determined by random variables uniformly distributed on the interval $(0,w_{i,max})$ at the start of the simulation for each agent. $\ln$ is the natural logarithm. $P_f$ is a fundamental value and is constant. $P^t$ is a mid-price (the average of the highest buy-order price and the lowest sell-order price) at $t$, and $\epsilon ^t _ j$ is determined by random variables from a normal distribution with average $0$ and variance $\sigma _ \epsilon$ at $t$. $\tau_j$ is independently determined by random variables uniformly distributed on the interval $(1,\tau _{max})$ at the start of the simulation for each agent\footnote{When $t< \tau _ j$, the second term of Eq. (\ref{eq1}) is zero.}.

The first term in Eq. (\ref{eq1}) represents a fundamental strategy: the agent expects a positive return when the market price is lower than the fundamental value, and vice versa. The second term represents a technical analysis strategy using a historical return: the agent expects a positive return when the historical market return is positive, and vice versa. The third term represents noise.

After the expected return has been determined, the expected price is
\begin{equation}
P^t_{e,j}= P^{t-1} \exp{(r^t_{e,j})}.
\end{equation}

Order prices are scattered around the expected price $P^t_{e,j}$ to replicate many waiting limit orders. An order price $P^t_{o,j}$ is 
\begin{equation}
P^t_{o,j}=P^t_{e,j}+P_d(2\rho ^t _j-1),
\end{equation}
where $\rho ^t_j$ is determined by random variables uniformly distributed on the interval $(0,1)$ at $t$ and $P_d$ is constant. This means that $P^t_{o,j}$ is determined by random variables uniformly distributed on the interval $(P^t_{e,j}-P_d, P^t_{e,j}+P_d)$. 

Whether the agent buys or sells is determined by the magnitude relationship between $P^t_{e,j}$ and $P^t_{o,j}$. Here\footnote{When $t<t_c$, to generate enough waiting orders, the agent places an order to buy one share when $P_f>P^t_{o,j}$, or to sell one share when $P_f<P^t_{o,j}$. 
\label{ft01}},

\begin{itemize}
\item[] when $P^t_{e,j}>P^t_{o,j}$, the agent places a buy order, and 
\item[] when $P^t_{e,j}<P^t_{o,j}$, the agent places a sell order. 
\end{itemize}

The remaining orders are canceled $t_c$ after the order time except when a circuit breaker is in operation.

\subsection{Erroneous orders} 
\label{s2.2}
We utilized the same model of erroneous orders as Mizuta et. al.\cite{mizuta2016ISAFM}. Erroneous orders start at time $t_{ms}$ and finish at time $t_{me}$ (see also Fig. \ref{p01} (left)). Within that period, with a constant probability $p_m$, each order of the agents is changed to one share sell at the highest buy-order price listed in the order book. The changed sell order is immediately executed matching the highest buy order. Increasing such erroneous sell orders is what makes market prices fall.

\subsection{Stop-loss orders} 
\label{s2.3}
The agent $j$ starts to place stop-loss orders when $P^t$ becomes under $P_f\exp{\epsilon ^{t=0} _j}-P_{l,j}$, where $P_{l,j}$ is independently determined by random variables uniformly distributed on the interval $(P_{lmin},P_{lmax})$ at the start of the simulation for each agent $j$. $t_{ls,j}$ is the start time of placing stop-loss orders for each agent. As shown on Fig. \ref{p01} (right), within the stop-loss period with a constant probability $p_l(t_{ls,j}+t_{l,j}-t)/t_{l,j}$, each order of an agent is changed to one share sell at the highest buy-order price, where $t_{l,j}$ is independently determined by random variables uniformly distributed on the interval $(t_{lmin},t_{lmax})$ at the start of the simulation for each agent $j$ and $p_l$ is constant.

Each agent estimates $P_f\exp{\epsilon ^{t=0} _j}$ as a fair price and has to re-estimates a fair price when market prices fall significantly below that price. The agent needs a long time for the re-estimation and places stop-loss orders during this process. In this study, because erroneous orders cause prices to fall significantly, the fair price is not changed. The agent learns this fact after the re-estimation. Therefore, the probability of a stop-loss order decreases as time progresses in this model, and the agent stops a stop-loss order after the re-estimation.

\subsection{Price limit and circuit breaker}
\label{s2.4}
We utilized the same model of a price limit as Mizuta et al.\cite{mizuta2016ISAFM}. In the case of adopting a price limit, as shown Fig. \ref{p02} (left), any order prices of sell under $P^{t-tr}-Pr$ (of buy above $P^{t-tr}+Pr$) are changed to $P^{t-tr}-Pr$ ($P^{t-tr}+Pr$), where $tr, Pr$ are constant parameters to determine the nature of the price limit. We simulated the model to investigate the nature of price formations when these parameters were changed. As Mizuta et al. \cite{mizuta2016ISAFM} demonstrated, the price limit prevents large price variations because agents cannot place orders far from $P^{t-tr}$ when the price limit is in place.

In the case of adopting a circuit breaker, as shown Fig. \ref{p02} (right), when $P_t$ reaches under $P^{t-tr}-Pr$ or above $P^{t-tr}+Pr$, the circuit breaker starts, and after starting, placing or cancelling orders are stopped during $tr_2$. Actually, the circuit breaker limits orders (unlike the price limit) and just waits for time to pass; however, the erroneous order period and stop-loss order period continue while the circuit breaker is activated. This leads to a decrease in the erroneous orders and stop-loss orders and, like the price limit, prevents large price variations.
We also introduced a price limit version two in which any order prices of sell under $P^{t-tr}-Pr$ or buy above $P^{t-tr}+Pr$ are canceled. In an actual financial market, investors will place orders exactly at the $P^{t-tr} \pm Pr$ with such a price limit because they know orders will be canceled outside the $P^{t-tr} \pm Pr$. Therefore, in real financial markets version two does not cause any changes and the result is exactly the same as the normal version. The reason we introduce version two is that we want to investigate the effect of multiple orders existing at $P^{t-tr} \pm Pr$, so we use agents that do not care whether their orders are cancelled even though this is unrealistic in actual financial markets.

\begin{figure}[t] 
\begin{center}
\includegraphics[scale=0.35]{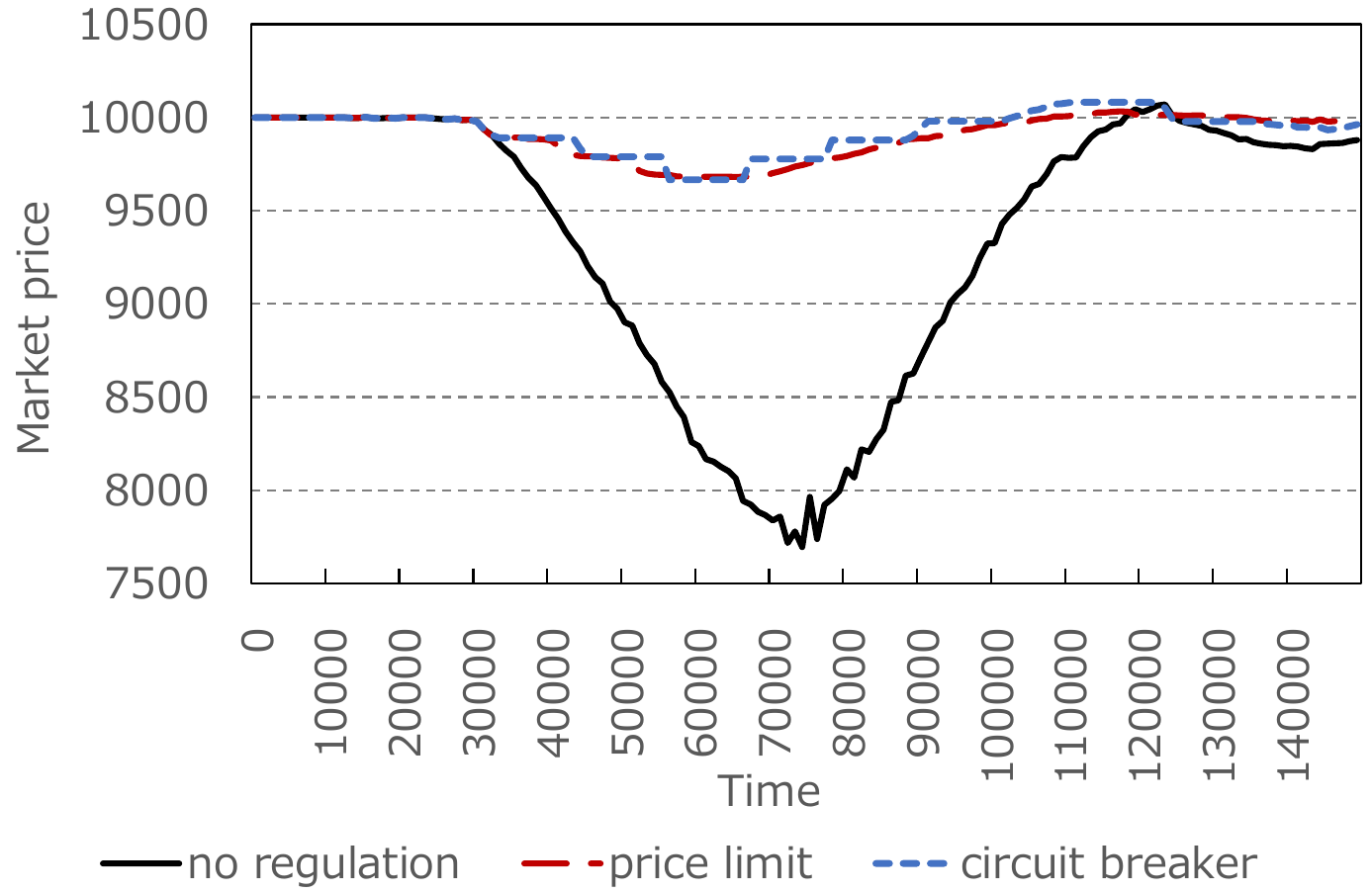}
\end{center}
\caption{Example of time evolution of market prices in cases with the price limit and the circuit breaker on $tr=10000 and Pr=100$, and with no regulation.}
\label{e01}
\end{figure}

\begin{figure}[t] 
\begin{center}
\includegraphics[scale=0.35]{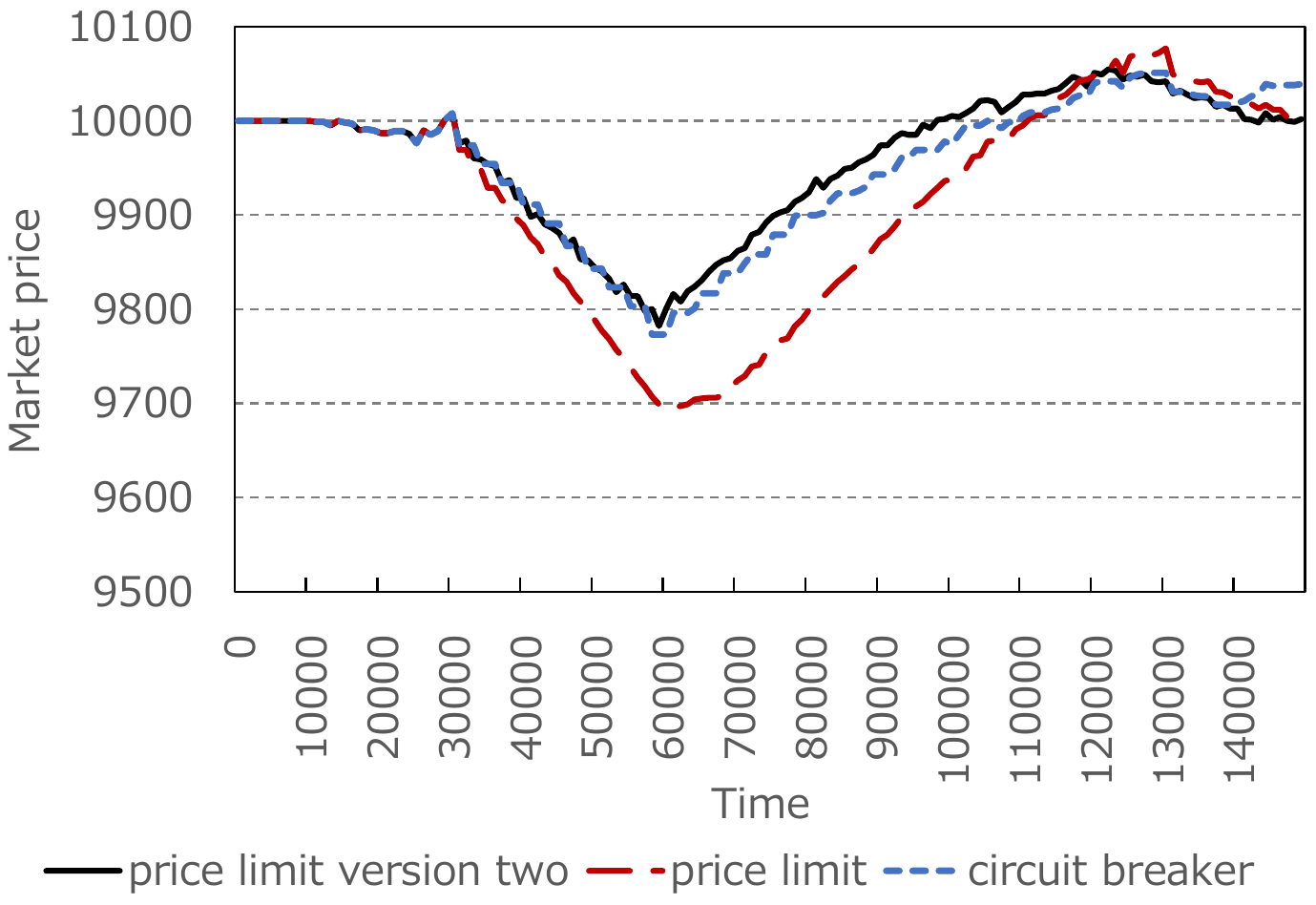}
\end{center}
\caption{Example of time evolution of market prices in cases with the price limit version two, the price limit, and the circuit breaker on $tr=2000 and Pr=20$.}
\label{e02}
\end{figure}

\begin{table}[t]
\caption{Averages of falling depth in cases with the price limit.}
\begin{center}
  \begin{tabular}{cr|rrrrrrr}
&&  \multicolumn{5}{c}{$tr$}   \\
&& 1000 & 2000 & 5000 & 10000 & 20000 \\ \hline
 & 10 & 298 & 157 & 73 & 45 & 33 \\ 
 & 20 & 577 & 302 & 134 & 79 & 52 \\
 & 50 & \cellcolor[gray]{0.90} 1227 & 721 & 313 & 167 & 110 \\
$Pr$ & 100 & \cellcolor[gray]{0.90} 2000 & \cellcolor[gray]{0.90} 1296 & 612 & 315 & 209 \\ 
 & 200 & \cellcolor[gray]{0.90} 2194 & \cellcolor[gray]{0.90} 2111 & 1140 & 615 & 408 \\
 & 500 & \cellcolor[gray]{0.90} 2054 & \cellcolor[gray]{0.90} 2054 & \cellcolor[gray]{0.90} 2053 & \cellcolor[gray]{0.90} 1419 & 906 \\
 & 1000 & \cellcolor[gray]{0.90} 2054 & \cellcolor[gray]{0.90} 2054 & \cellcolor[gray]{0.90} 2054 & \cellcolor[gray]{0.90} 2054 & \cellcolor[gray]{0.90} 1433  
\end{tabular}
\label{t01}
\end{center}
\end{table}

\begin{table}[t]
\caption{Averages of falling depth in cases with the circuit breaker.}
\begin{center}
  \begin{tabular}{cr|rrrrrrr}
&&  \multicolumn{5}{c}{$tr$}   \\
&& 1000 & 2000 & 5000 & 10000 & 20000 \\ \hline
 & 10 & 152 & 103 & 57 & 34 & 22 \\
 & 20 & 297 & 207 & 107 & 58 & 43 \\
 & 50 & \cellcolor[gray]{0.90} 702 & 487 & 266 & 163 & 105 \\
 $Pr$ & 100 & \cellcolor[gray]{0.90} 1222 & \cellcolor[gray]{0.90} 869 & 520 & 315 & 208 \\
 & 200 & \cellcolor[gray]{0.90} 1954 & \cellcolor[gray]{0.90} 1522 & 821 & 614 & 407 \\
 & 500 & \cellcolor[gray]{0.90} 2054 & \cellcolor[gray]{0.90} 2054 & \cellcolor[gray]{0.90} 2038 & \cellcolor[gray]{0.90} 1061 & 503 \\
 & 1000 & \cellcolor[gray]{0.90} 2054 & \cellcolor[gray]{0.90} 2054 & \cellcolor[gray]{0.90} 2054 & \cellcolor[gray]{0.90} 2054 & \cellcolor[gray]{0.90} 1006 
\end{tabular}
\label{t02}
\end{center}
\end{table}

\begin{table}[t]
\caption{Differences between averages of falling depth with the price limit and with the circuit breaker.}
\begin{center}
  \begin{tabular}{cr|rrrrrrr}
&&  \multicolumn{5}{c}{$tr$}   \\
&& 1000 & 2000 & 5000 & 10000 & 20000 \\ \hline
 & 10 & 146 & 54 & 16 & 12 & 11 \\
 & 20 & 280 & 95 & 27 & 21 & 9 \\
 & 50 & \cellcolor[gray]{0.90}525 & 234 & 47 & 4 & 5 \\
 $Pr$ & 100 & \cellcolor[gray]{0.90}778 & \cellcolor[gray]{0.90}427 & 92 & 0 & 1 \\
 & 200 & \cellcolor[gray]{0.90}240 & \cellcolor[gray]{0.90}589 & 319 & 1 & 1 \\
 & 500 & \cellcolor[gray]{0.90}0 & \cellcolor[gray]{0.90}0 & \cellcolor[gray]{0.90}15 & \cellcolor[gray]{0.90}359 & 403 \\
 & 1000 & \cellcolor[gray]{0.90}0 & \cellcolor[gray]{0.90}0 & \cellcolor[gray]{0.90}0 & \cellcolor[gray]{0.90}0 & \cellcolor[gray]{0.90}427 
\end{tabular}
\label{t03}
\end{center}
\end{table}

\begin{table}[t]
\caption{Differences between averages of falling depth with the price limit version two and with the circuit breaker.}
\begin{center}
  \begin{tabular}{cr|rrrrrrr}
&&  \multicolumn{5}{c}{$tr$}   \\
&& 1000 & 2000 & 5000 & 10000 & 20000 \\ \hline
 & 10 & -8 & -17 & -13 & -5 & -1 \\
 & 20 & 56 & 3 & -4 & 2 & -3 \\
 & 50 & \cellcolor[gray]{0.90}212 & 83 & 16 & -14 & -6 \\
 $Pr$ & 100 & \cellcolor[gray]{0.90}418 & \cellcolor[gray]{0.90}247 & 38 & -17 & -10 \\
 & 200 & \cellcolor[gray]{0.90}69 & \cellcolor[gray]{0.90}413 & 231 & -23 & -10 \\
 & 500 & \cellcolor[gray]{0.90}0 & \cellcolor[gray]{0.90}0 & \cellcolor[gray]{0.90}15 & \cellcolor[gray]{0.90}318 & 392 \\
 & 1000 & \cellcolor[gray]{0.90}0 & \cellcolor[gray]{0.90}0 & \cellcolor[gray]{0.90}0 & \cellcolor[gray]{0.90}0 & \cellcolor[gray]{0.90}417
\end{tabular}
\label{t04}
\end{center}
\end{table}

\begin{figure}[t] 
\begin{center}
\includegraphics[scale=0.45]{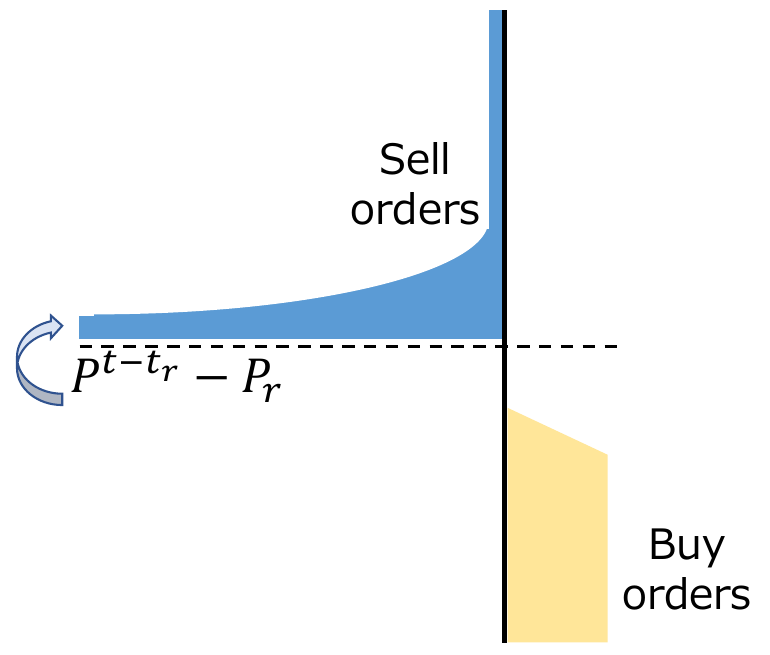}
\end{center}
\caption{An illustration of multiple sell orders accumulated around the lower limit price.}
\label{p03}
\end{figure}

\begin{table*}
\caption{The order books at $t=t_{me}=60000$ in the example from Fig. \ref{e02}.}
\begin{center}
\begin{tabular}{lll}
\begin{tabular}{c|c|c}
\multicolumn{3}{c}{Price limit} \\ \multicolumn{3}{c}{} \\
sell shares & price & buy shares \\ \hline
272 &9780-9800 & \\	
290 &9760-9780 & \\
279 &9740-9760 & \\ 
278 &9720-9740 & \\
136 &9700-9720 & \\
& 9680-9700 & 49 \\
& 9660-9680 & 69 \\
& 9640-9660 & 83 \\
& 9620-9640 & 80 \\
& 9600-9620 & 67 \\
\end{tabular}

&

\begin{tabular}{c|c|c}
\multicolumn{3}{c}{Circuit breaker} \\ \multicolumn{3}{c}{} \\
sell shares & price & buy shares \\ \hline
6 &9860-9880 & \\	
9 &9840-9860 & \\
8 &9820-9840 & \\ 
1 &9800-9820 & \\
1 &9780-9800 & \\
 & 9760-9780 & 17 \\
 & 9740-9660 & 44 \\
 & 9720-9740 & 44 \\
 & 9700-9720 & 64 \\
 & 9680-9700 & 54 \\
\end{tabular}

&

\begin{tabular}{c|c|c}
\multicolumn{3}{c}{Price limit version two} \\ \multicolumn{3}{c}{} \\
sell shares & price & buy shares \\ \hline
82 &9880-9900 & \\	
85 &9860-9880 & \\
55 &9840-9860 & \\ 
25 &9820-9840 & \\
3 &9800-9820 & \\
 & 9700-9800 & 4 \\
 & 9720-9780 & 74 \\
 & 9740-9760 & 72 \\
 & 9720-9740 & 86 \\
 & 9700-9720 & 88 \\
\end{tabular}

\end{tabular}
\end{center}
\label{t05}
\end{table*}

\section{Simulation results}
\label{s4}

In this study, we determined and validated the parameters of model to replicate the fat-tail and volatility clustering that are stably observed as stylized facts for any asset and in any period same as Mizuta et al. \cite{mizuta2016ISAFM}. We did not dare to replicate other stylized facts that are unstably observed because, as we mentioned, the simplicity of the model is very important for this study because unnecessary replication of macro phenomena leads to models that are overfitted and too complex. More details, see the Appendix \ref{s5.2} and \ref{s5.3}.

Then, we set $\delta P=0.01 and P_f=10000$, and for the agents, $n=1000, w_{1,max}=1, w_{2,max}=10, w_{3,max}=1, \tau _ {max}=10000, \sigma _ \epsilon = 0.03, P_d= 1000, t_c=10000$, and $P_{f}=10000$. For the erroneous orders, we set $t_{ms}=30000, t_{me}=60000, and p_m=0.15$, and for stop-loss orders, $P_{lmin}=1000, P_{lmax}=3000, t_{lmin}=10000, t_{lmax}=100000, and p_l=0.35$. The simulations ran to $t=t_e=150000$. 

The price limit and the circuit breaker have the same parameters ($tr$ and $Pr$, respectively). We simulated $35 (=5 \times 7)$ cases for $tr=1000, 2000, 5000, 10000, and 20000$, and for $Pr=10, 20, 50, 100, 200, 500, and 1000$, where the other parameters were fixed and used the same random number table. We then simulated these runs 100 times, changing the random number table each time, and counted the averages of falling depth ($P_f-$lowest $P^t$; see also Fig. \ref{p01} (left)).

We used the above ranges of $tr$ and $Pr$ so as to satisfy the conditions shown by Eqs. (\ref{eqj1})--(\ref{eqj4}) in Section \ref{s1}. Specifically, we set $Pr$ to satisfy the conditions in Eqs. (\ref{eqj2}) and (\ref{eqj4}) and $tr$ to satisfy the condition in Eq. (\ref{eqj3}). The results are shown in Tables \ref{t01}--\ref{t04}, where the gray shading indicates Eq. (\ref{eqj1}) is not satisfied.

Figure \ref{e01} shows an example of the time evolution of market prices (mid prices) in the cases with the price limit and the circuit breaker on $tr=10000, Pr=100$, and without either. Note that while the example is just one out of 100 simulation runs, the market prices of all cases showed a very similar shape. In the case without the regulations, market prices fell sharply and the falling continued by around $t=72000$ after $t=t_{me}=60000$ when the erroneous orders were stopped, which means that an overshoot occurred. This result is consistent with that reported by Mizuta et al. \cite{mizuta2016ISAFM}. In the cases with the price limit and the circuit breaker, market prices did not significantly fall and they started to increase after $t=t_{me}=60000$ when the erroneous orders were stopped. This means that an overshoot did not occur.

Tables \ref{t01} and \ref{t02} list the averages of falling depth ($P_f-$lowest $P^t$; see also Fig. \ref{p01} (left)) for various $tr, Pr$ in the cases with the price limit and the circuit breaker, where gray shading indicates Eq. (\ref{eqj1}) is not satisfied. The falling speed is calculated around $0.052$ on the basis of Fig. \ref{e01}. As we can see, market prices fell significantly when Eq. (\ref{eqj1}) was not satisfied, but this could be prevented when Eq. (\ref{eqj1}) was satisfied. At $tr \ge 10000$, the price limit and the circuit breaker showed almost the same effectiveness to prevent the falling, but at $tr<10000$, the circuit breaker was more effective. To clarify this, Table \ref{t03} shows the differences in falling depth averages between the price limit and the circuit breaker.

Next we examine the mechanism underlying these behaviors. In the price limit case, sharply falling prices cause many sell orders to be accumulated around the lower limit price, $p^{t-tr}-Pr$. The threshold, $tr=10000$, is exactly the same as the cancel period, $t_c=10000$. Therefore, when $tr<tc$, $p^{t-tr}-Pr$ is changed before the accumulated sell orders are cancelled, which leads to the accumulation of more sell orders of various prices, as shown in Fig. \ref{p03}. This accumulation then acts like a wall against buy orders, which prevents rising prices . Therefore, it is more difficult for the price limit to prevent price falling when $tr<tc$ than when $tr>tc$.  

Actually, in the case with the price limit version two, which does not accumulate sell orders around the lower limit price, falling is prevented more or less as effectively as the circuit breaker when $tr<tc$, as Table \ref{t04} indicates. Figure \ref{e02} shows an example of the time evolution of market prices (mid prices) in the cases with the price limit version two, the price limit, and the circuit breaker on $tr=2000, Pr=20$. As we can see, the time evolution market prices with the price limit version two were almost the same as those with the circuit breaker.

Table \ref{t05} shows the order book at $t=t_{me}=60000$ for the example in Fig. \ref{e02}. The numbers are aggregated orders for price ranges in increments of 20. In the case with the price limit, many sell orders were accumulated and they prevented the prices from rising.

To summarize, the price limit and the circuit breaker basically prevent falling prices with the same level of effectiveness when the $Pr and tr$ parameters are the same. However, the price limit is less effective when $tr<tc$. In the case with the price limit, many sell orders are accumulated around the lower limit price. When $tr<tc$, $p^{t-tr}-Pr$ is changed before the accumulated sell orders are cancelled, which leads to the accumulation of more sell orders of various prices. This accumulation acts like a wall against buy orders, and the wall is what prevents the prices from rising. 

\section{Summary}
\label{s5}

In this study, we expanded on Mizuta et al.'s artificial market model\cite{mizuta2016ISAFM} by adding stop loss behavior of the agents and a circuit breaker and then used it to investigate whether the price limit or the circuit breaker is more effective to prevent falling market prices.
Our findings showed that the price limit and the circuit breaker are basically equally effective when the limit price range ($Pr$) and limit time range ($tr$) parameters are the same. However, the price limit is less effective when $tr$ is smaller than the cancel time range ($tc$). In the case with the price limit, many sell orders are accumulated around the lower limit price. When $tr<tc$, the lower limit price is changed before the accumulated sell orders are cancelled, which leads to the accumulation of more sell orders of various prices. This accumulation then acts like a wall against buy orders, which prevents the prices from rising. 

Caution should be taken in the sense that these results pertain only to a limited situation. Specifically, the fact that the circuit breaker is better than the price limit should be adapted only in the case where the reason for falling prices is erroneous orders and where individual stocks are regulated. In cases where other reasons make the market prices fall (for example, large variations in the fundamental price) or in cases where individual stocks are not regulated (for example, stock indexes are regulated), the results might be different. In fact, in real financial markets, market prices frequently fall for reasons other than erroneous orders, and there are many cases where a stock index is regulated by circuit breakers while individual stocks are not regulated. Wang et. al. \cite{WANG2022106533} indicated that if a stock index is regulated and the index closes to the limit price of circuit breaker, some traders are in a hurry to sell and their sell orders make the index fall, which means that sometimes a circuit breaker induces a sharp fall rather than preventing it. In the current work, we did not examine such behavior of traders, so this and the other caveats mentioned above will be the focus of our future work.

\section{Appendix} \label{s5}

\subsection{Brief history and contributions of Artificial market models} \label{s5.1}
An artificial market model is an agent-based model for a financial market. There are thorough great reviews on such model \cite{LEBARON20061187, doi:10.1080/14697688.2010.539249, chen2009agent, Cristelli2014, 7850016, mizuta2016SSRNrev, mizuta2019arxiv, mizuta2022aruka, segovia2022bibliometric}. Numerous significant artificial market models \cite{TAKAYASU1992127, izumi1996artificial, arthur1997economy, lux1999scaling} have been developed since their first appearance in the 1990s. Because of their generality, they are often qualitative. In particular, artificial markets have been used to investigate the mechanism by which stylized facts\footnote{A stylized fact is a term used in economics to refer to empirical findings that are so consistent (for example, across a wide range of instruments, markets and time periods) that they are accepted as truth\cite{Sewell2006}.} (fat-tails, volatility-clustering, and so on) emerge \cite{TAKAYASU1992127,lux1999scaling}. 

Here, let us briefly examine the nature of financial market phenomena such as bubbles and crashes, as described in \cite{izumi1996artificial,arthur1997economy}. It is said that micro-macro feedback loops have played very important roles in bubbles and crashes, and artificial market models can directly treat such loops. A number of projects have built generic artificial market models, such as the U-mart project in Japan in the 2000s. (Kita et al. provide a comprehensive review of the U-mart project\cite{kitarealistic2016}.) These projects have helped to explain the nature of financial market phenomena and the mechanism by which stylized facts emerge.

Artificial market models, however, have rarely been used to investigate the rules and regulations of financial markets. After the bankruptcy of Lehman Brothers in 2008, some researchers argued that traditional economics had not found ways to design markets that work well but suggested that an artificial market model could do so. Indeed, in Nature, Farmer and Foley \cite{farmer2009economy} explained that ``such (agent based) economic models should be able to provide an alternative tool to give insight into how government policies could affect the broad characteristics of economic performance, by quantitatively exploring how the economy is likely to react under different scenarios.'' Richard Bookstaber, an expert on risk management  who has worked for investment banks and hedge funds, wrote a book \cite{Bookstaber2017} that ``provides a nontechnical introduction to agent-based modeling, an alternative to neoclassical economics that shows great promise in predicting crises, averting them, and helping us recover from them.'' In 2010, Jean-Claude Trichet, then President of the European Central Bank (ECB) \cite{ECB2010}, stated that ``agent-based modelling dispenses with the optimization assumption and allows for more complex interactions between agents. Such approaches are worthy of our attention.''

Financial regulators and exchanges, who decide rules and regulations, are especially interested in using artificial market models to design a market that works well. Indeed, the Japan Exchange Group (JPX), which is the parent company of the Tokyo Stock Exchange, has published 40 JPX working papers, including 12 on using artificial market models, as of December 2022\footnote{\url{https://www.jpx.co.jp/english/corporate/research-study/working-paper/index.html}}.


Mizuta \cite{mizuta2016SSRNrev} reviewed other previous agent-based models for designing a financial market that works well.

\subsection{Basic concept for constructing a model} \label{s5.2}

An artificial market model, which is an agent-based model for financial markets, can be used to investigate situations that have never occurred, handle regulation changes that have never been made, and isolate the pure contribution of these changes to price formation and liquidity\cite{mizuta2019arxiv, mizuta2022aruka}. These are the advantages of an artificial market simulation. However, the outputs of this simulation would not be accurate or be credible forecasts of the actual future. The simulation needs to reveal possible mechanisms that affect price formation through many simulation runs, e.g., searching for parameters or purely comparing the before and after states of changes. The possible mechanisms revealed by these runs provide new intelligence and insight into the effects of the changes in price formations in actual financial markets. Other methods of study, e.g., empirical studies, would not reveal such possible mechanisms.

Artificial markets should replicate the macro phenomena that exist generally for any asset at any time. Price variation, which is a kind of macro phenomenon, is not explicitly modeled in artificial markets. Only micro processes, agents (general investors), and price determination mechanisms (financial exchanges) are explicitly modeled. Macro phenomena emerge as the outcome of interactions from micro processes. Therefore, the simulation outputs should replicate existing macro phenomena to generally prove that simulation models are probable in actual markets.

However, it is not the primary purpose for an artificial market to replicate specific macro phenomena only for a specific asset or period. Unnecessary replication of macro phenomena leads to models that are overfitted and too complex. Such models would prevent us from understanding and discovering mechanisms that affect price formation because the number of related factors would increase. In addition, artificial market models that are too complex are often criticized because they are very difficult to evaluate\cite{chen2009agent}. A model that is too complex not only would prevent us from understanding mechanisms but also could output arbitrary results by overfitting too many parameters. It is more difficult for simpler models to obtain arbitrary results, and these models are easier to evaluate.

Therefore, we constructed an artificial market model that is as simple as possible and does not intentionally implement agents to cover all the investors who would exist in actual financial markets.

Such simplicity is very important not only for artificial market models but also generally for agent-based models. Gilbert argued that there are three types for agent-based models, an abstract model, a middle-range model and a facsimile model\cite{gilbert2008agent}. Gilbert said "the aim of abstract models is to demonstrate some basic social process that may lie behind many area of social life." Axelrod mentioned that to understand mechanism abstract models should be as simple as possible because needless complex model preventing to understand the mechanism, and called the principle as KISS (keep it simple stupid)\cite{10.2307/j.ctt7s951}. For example, Thomas Schelling, who received the Nobel Prize in economics, used an agent-based model to discuss the mechanism of racial segregation. The model was built very simply compared with an actual town to focus on the mechanism\cite{Schelling2006}. While it was not able to predict the segregation situation in the actual town, it was able to explain the mechanism of segregation as a phenomenon. Of course, the model was not calibrated to empirical data because of not leading to make it more complex to prevent understanding. For such abstract models, we should be cautious complex modeling by calibrations using empirical date, and this is a fundamental for modeling\cite{Weisberg2012}. Indeed, many artificial market models were validated by replicating fat-tails and volatility-clustering that are very famous stylized facts in financial markets\cite{LEBARON20061187, chen2009agent, mizuta2019arxiv, mizuta2022aruka}. On the other hand, the aim of facsimile models is to replicate some specific situation and "predict" future exactly. So, facsimile models are needed to be calibrated by empirical data at least and should replicate existing social phenomena exactly.

As Michael Weisberg mentioned\cite{Weisberg2012}, ``Modeling, (is) the indirect study of real-world systems via the construction and analysis of models.'' ``Modeling is not always aimed at purely veridical representation. Rather, they worked hard to identify the features of these systems that were most salient to their investigations.'' Therefore, effective models are different depending on the phenomena they focus on. Thus, our model is effective only for the purpose of this study and not for others. The aim of our study is to understand how important properties (behaviors, algorithms) affect macro phenomena and play a role in the financial system rather than representing actual financial markets precisely.

The aforementioned discussion holds not only for artificial markets but also for agent-based models used in fields other than financial markets. For example, Thomas Schelling, who received the Nobel Prize in economics, used an agent-based model to discuss the mechanism of racial segregation. The model was built very simply compared with an actual town to focus on the mechanism\cite{Schelling2006}. While it was not able to predict the segregation situation in the actual town, it was able to explain the mechanism of segregation as a phenomenon.


Michael Weisberg studied what mathematical and simulation models are in the first place and cited the example of a map\cite{Weisberg2012}. Needless to say, a map models geographical features on the way to a destination. With a simple map, we can easily understand the way to the destination. However, while a satellite photo replicates actual geographical features very well, we cannot easily find the way to the destination.

The title page of Michael Weisberg's book\cite{Weisberg2012} cited a passage from a short story by Jorge Borges\cite{Borges}, ``In time, those Unconscionable Maps no longer satisfied, and the Cartographers Guilds struck a Map of the Empire whose size was that of the Empire, and which coincided point for point with it...In the Deserts of the West, still today, there are Tattered Ruins of that Map, inhabited by Animals and Beggars.'' The story in which a map was enlarged to the same size as the real Empire to become the most detailed of any map is an analogy to that too detailed a model is not useful. This story give us one of the most important lessons for when we build and use any model.

\begin{table}[t]
\caption{Stylized facts without the erroneous orders}
\begin{center}
 \begin{tabular}{ccr}
 \multicolumn{2}{c}{kurtosis or returns} & $4.32$ \\ \hline
 & lag & \\
 & 1 & $0.130$ \\
 autocorrelation & 2 & $0.081$ \\
 coefficient for & 3 & $0.065$ \\
 square returns & 4 & $0.054$ \\
 & 5 & $0.045$ 
 \end{tabular}
\label{t0}
\end{center}
\end{table}

\subsection{Validation of the model} \label{s5.3}
In many previous artificial market studies, the models were validated to determine whether they could explain stylized facts, such as a fat-tail or volatility clustering\cite{LEBARON20061187, chen2009agent, mizuta2019arxiv, mizuta2022aruka}. A fat-tail means that the kurtosis of price returns is positive. Volatility clustering means that square returns have a positive autocorrelation, which slowly decays as its lag becomes longer.

Many empirical studies, e.g., that of Sewell\cite{Sewell2006}, have shown that both stylized facts (fat-tail and volatility clustering) exist statistically in almost all financial markets. Conversely, they also have shown that only the fat-tail and volatility clustering are stably observed for any asset and in any period because financial markets are generally unstable. This leads to the conclusion that an artificial market should replicate macro phenomena that exist generally for any asset at any time, fat-tails, and volatility clustering. Other stylized facts should be replicate only when a purpose of study relates them to prevent to make a model needless complex. In the case of this study, our model should replicate only universally established stylized facts about time evolution of market prices which are fat-tails and volatility clustering to prevent to make a model needless complex because we focus generally and universally impacts to market prices by the rules. We did not dare to replicate other stylized facts that are unstably observed because, as we mentioned, the simplicity of the model is very important for this study because unnecessary replication of macro phenomena leads to models that are overfitted and too complex. This is an example of how empirical studies can help an artificial market model.

The kurtosis of price returns and the autocorrelation of square returns are stably and significantly positive, but the magnitudes of these values are unstable and very different depending on the asset and/or period. Very broad magnitudes of about $1 \sim 100$ and about $0 \sim 0.2$, respectively, have been observed\cite{Sewell2006}.

For the aforementioned reasons, an artificial market model should replicate these values as significantly positive and within a reasonable range. It is not essential for the model to replicate specific values of stylized facts because the values of these facts are unstable in actual financial markets.

Table \ref{t0} lists the statistics showing the stylized facts, kurtosis of price returns for $100$ tick times ($\ln(P^t/P^{t-100})$), and autocorrelation coefficient for square returns for $100$ tick times without the erroneous orders. Note that without the erroneous orders no stop loss is happened and no regulation is triggered. This shows that this model replicated the statistical characteristics, fat-tails, and volatility clustering observed in real financial markets.

\bibliographystyle{IEEEtran}
\bibliography{ref}

\end{document}